\documentclass[aps,
prx,
superscriptaddress,
amsmath,
amssymb,
reprint,
floatfix]
{revtex4-2}

\usepackage[utf8]{inputenc}
\usepackage{graphicx}
\usepackage{color}
\usepackage{qcircuit}
\usepackage{braket}
\usepackage{url}
\usepackage{bm}
\usepackage[version=4]{mhchem}
\usepackage{mathtools}
\usepackage{siunitx}
\usepackage[breaklinks=true]{hyperref}
\newcommand{\mr}[1]{\mathrm{#1}}

\newcommand{\tr}{\mathrm{Tr}}

\date{\today}

\begin{document}
\title{Penalty methods for variational quantum eigensolver}

\author{Kohdai Kuroiwa}
\email{kkuroiwa@uwaterloo.ca}
\affiliation{Institute for Quantum Computing and Department of Physics and Astronomy, University of Waterloo, 200 University Avenue West, Waterloo, Ontario, Canada, N2L 3G1}

\author{Yuya O. Nakagawa}
\email{nakagawa@qunasys.com}
\affiliation{QunaSys Inc., Aqua Hakusan Building 9F, 1-13-7 Hakusan, Bunkyo, Tokyo 113-0001, Japan}

\begin{abstract}
The variational quantum eigensolver (VQE) is a promising algorithm to compute eigenstates and eigenenergies of a given quantum system that can be performed on a near-term quantum computer.
Obtaining eigenstates and eigenenergies in a specific symmetry sector of the system is often necessary for practical applications of the VQE in various fields ranging from high energy physics to quantum chemistry. 
It is common to add a penalty term in the cost function of the VQE to calculate such a symmetry-resolving energy spectrum; however, systematic analysis of the effect of the penalty term has been lacking, and the use of the penalty term in the VQE has not been justified rigorously.
In this work, we investigate two major types of penalty terms for the VQE that were proposed in previous studies.
We show that a penalty term of one of the two types works properly in that eigenstates obtained by the VQE with the penalty term reside in the desired symmetry sector.
We further give a convenient formula to determine the magnitude of the penalty term, which may lead to faster convergence of the VQE.
Meanwhile, we prove that the other type of penalty term does not work for obtaining the target state with the desired symmetry in a rigorous sense and even gives completely wrong results in some cases.
We finally provide numerical simulations to validate our analysis.
Our results apply to general quantum systems and lay the theoretical foundation for the use of the VQE with the penalty terms to obtain the symmetry-resolving energy spectrum of the system, which fuels the application of a near-term quantum computer.
\end{abstract}

\maketitle
\section{\label{sec:intro} Introduction}
Quantum computers have been attracting attention because they are expected to solve some classes of computation problems remarkably faster than classical computers~\cite{grover1996fast,shor1999polynomial,Lloyd_2014,Harrow_2009}.
Quantum computers that are expected to be realized in the near future are called noisy intermediate-scale quantum (NISQ) devices, which contain a few hundred qubits and do not perform error correction on their qubits~\cite{Preskill2018}.
Although NISQ devices are not suitable to perform such algorithms~\cite{grover1996fast,shor1999polynomial,Lloyd_2014,Harrow_2009} that practically require error correction with numerous qubits, simulating quantum many-body systems is one of the most promising ways to utilize NISQ devices. In particular, the variational quantum eigensolver (VQE)~\cite{peruzzo2014variational}, which was already performed on a real NISQ device~\cite{peruzzo2014variational,kandala2017hardware,colless2018computation}, computes the ground-state energy and the ground state of a given quantum system, for example, molecules~\cite{mcardle2018quantum,Cao2018}.
In the VQE, we prepare a parametrized state, an \textit{ansatz} state, on a quantum computer and minimize the expectation value of energy for this state by updating the parameters. 
Recently, variations of the VQE have been developed to compute not only the ground state and its energy but also excited states/energies~\cite{McClean2017hybrid,nakanishi2018subspace,Parrish2019,colless2018computation,jones2019variational,higgott2019variational,ollitrault2019quantum}.

In many cases appearing in various fields ranging from high-energy physics to condensed matter physics, a Hamiltonian we encounter possesses some symmetry. 
It is then of great importance to obtain the energy spectrum of the Hamiltonian for each symmetry sector (eigenvalue of the symmetry operation) to analyze the properties of the system.
For example, such a symmetry-resolving spectrum is important to predict photochemical reactions in quantum chemistry~\cite{Turro2009}.
It can also tell the nature of the spontaneous symmetry breaking in condensed matter physics~\cite{Anderson1952}.
When we want to use the VQE algorithm to compute eigenvalues/eigenstates of a given Hamiltonian that reside in a target sector of the symmetry, there are several options.
The first one is to use a special ansatz state in the VQE so that the ansatz state always belongs to the desired sector of the symmetry~\cite{Gard2019}, but the construction of such an ansatz strongly depends on the heuristic knowledge of the symmetry and is not always readily available for generic symmetries.
The second one is to ``taper off" the qubits by finding a unitary transformation in which eigenvalues of the symmetry are encoded into specific qubits~\cite{bravyi2017tapering}.
This technique is more general compared with the first one, but the sector (eigenvalue) of the symmetry can be designated only in modulo 2. 
The third option, which is the most general and which we study in this work, is to include penalty terms in the VQE algorithm to aim to guarantee that the resulting state of the VQE has the desired symmetry~\cite{mcclean2016theory,Rubin_2018,greenediniz2019generalized,higgott2019variational,Ilya2019}. 
We refer to such a strategy, or the VQE with penalty terms, as the constrained VQE in this work following Ref.~\cite{Ilya2019}.

In previous work, two types of penalty terms (detailed definitions are found in Eqs.~\eqref{eq:cost_func1} and \eqref{eq:cost_func2}) have been suggested to perform the constrained VQE, \textit{i.e.}, to compute eigenstates possessing the desired symmetry.
However, there are serious problems in such penalty terms that can hinder practical applications of the constrained VQE.
First, systematic analysis of the coefficients of penalty terms,  \textit{penalty coefficients}, has been lacking.
The coefficients stipulate the magnitude of the penalty for the ansatz state that deviates from the desired symmetry sector.
Small coefficients are preferable for the fast convergence of the VQE, but when they are too small, the resulting state of the VQE does not necessarily have the desired symmetry.
Currently, the coefficients are heuristically chosen, and there is no guarantee that one can obtain the state with the desired symmetry as a result of the VQE.
Second, the performance of the two types of penalty terms has not been explored.
Evidence for which one is superior to the other greatly helps when we utilize the constrained VQE in practical problems.

In this work, we theoretically study the two types of penalty terms and report several crucial findings to exploit the constrained VQE in actual calculations.
For one type of penalty term (Eq.~\eqref{eq:cost_func1}), we provide a convenient formula to determine the penalty coefficients that can rigorously ensure the result of the constrained VQE has correct symmetry.
On the other hand, remarkably, we prove that the other type of penalty term (Eq.~\eqref{eq:cost_func2}) does not work for {\it any} finite value of the coefficients in a rigorous sense.
We show that it even yields a completely wrong answer in some cases.
In addition, we also briefly analyze the effects of noise on the performances of these two penalty terms. 
We reveal that while the first one is robust to the effects of noise, the other one is susceptible to error. 
To validate our theoretical analysis, we provide numerical experiments simulating the constrained VQE.
Our findings resolve the problems of the constrained VQE and pave the way for the further applications of it.
The analysis of energy spectra with resolving their symmetry is ubiquitous in the study of modern quantum physics; our results have a broad impact on the use of near-term quantum computers (or NISQ devices) in various fields.

The rest of this paper is organized as follows. 
In Sec.~\ref{sec:prelim}, we review several basic concepts underlying the constrained VQE. 
We provide our main results in Sec.~\ref{sec:results}.
We analyze the two types of cost functions in the constrained VQE and derive several implications for both of them. 
In Sec.~\ref{sec:simulation}, we perform numerical simulations to validate our theoretical results in Sec.~\ref{sec:results}. 
In Sec.~\ref{sec:conclusion}, we summarize our results and give a conclusion.

\section{\label{sec:prelim} Preliminaries}
In this section, we review several concepts necessary to understand our main results.
In Sec.~\ref{subsec:VQE}, we briefly explain the VQE algorithm.  
In Sec.~\ref{subsec:VQD}, we review a variant of the VQE algorithm to obtain excited states, which is called the variational quantum deflation (VQD) algorithm. 
Finally, in Sec.~\ref{subsec:constraints}, we introduce the constrained VQE/VQD, or the VQE/VQD with penalty terms. 

\subsection{\label{subsec:VQE} Variational Quantum Eigensolver (VQE)}
The VQE~\cite{peruzzo2014variational} is a quantum-classical hybrid algorithm to compute the ground state and its energy of a given system and is considered to be executable on NISQ devices. 
Suppose that we have a quantum system with a Hamiltonian $\hat{H}$. We prepare an ansatz state of $n$ qubits,
\begin{equation}
    \ket{\psi(\bm{\theta)}} \coloneqq V(\bm{\theta})\ket{\psi_{0}}, \label{eq: ansatz state}
\end{equation}
using an ansatz circuit (unitary operation)
\begin{equation}
    V(\bm{\theta}) \coloneqq V_l(\theta_l)\cdots V_2(\theta_2)V_1(\theta_1), 
\end{equation}
where $\bm{\theta} \coloneqq (\theta_1,\ldots,\theta_l)$ are real-valued parameters, $V_i(\theta_i)$ is a parametrized quantum gate, and $\ket{\psi_0}$ is some reference state.
In the VQE, we optimize these parameters so that the expectation value of the energy,
\begin{equation}
    L_{\mr{VQE}}(\bm{\theta}) \coloneqq \bra{\psi(\bm{\theta})}\hat{H}\ket{\psi(\bm{\theta})},
\end{equation}
is minimized with respect to $\bm{\theta}$.
Then, letting $\bm{\theta}^*$ be the optimum, we can regard $L_{\mr{VQE}}(\bm{\theta}^*)$ as a nice approximation of the ground-state energy and $\ket{\psi(\bm{\theta}^*)}$ as that of the ground state due to the variational principle.
During the process of the algorithm, the preparation of the ansatz state $\ket{\psi(\bm{\theta})}$ and the measurements necessary to compute $L_{\mr{VQE}}(\bm{\theta})$ are performed by a quantum computer while the optimization is completed solely by a classical computer.
This separation of roles between classical and quantum computers alleviates the hardware requirement for quantum computers and makes the algorithm possible to run even on NISQ devices.

\subsection{\label{subsec:VQD} Variational Quantum Deflation (VQD)}
The VQE has been extended to compute excited states~\cite{McClean2017hybrid,nakanishi2018subspace,Parrish2019,colless2018computation,jones2019variational,higgott2019variational,ollitrault2019quantum}.
Among those extensions, the variational quantum deflation (VQD)~\cite{higgott2019variational} is shown to be possibly the most efficient~\cite{ibe2020calculating} compared with the two major ones, subspace-search VQE (SSVQE)~\cite{nakanishi2018subspace} and the multistate contracted VQE (MCVQE)~\cite{Parrish2019}.
In this work, we focus on the VQD and review it here.
We note that our analysis in Sec.~\ref{sec:results} also applies to the SSVQE and MCVQE (see the second-to-last paragraph of Sec.~\ref{subsec: construction_mu}).

Suppose that our goal is to obtain the $k$th excited-state energy of $\hat{H}$.
The VQD algorithm assumes that we already obtain the eigenstates up to the $(k-1)$th as $\ket{\psi_{i}} (i = 0,1,\ldots,k-1)$.
The cost function to be minimized in the VQD algorithm is
\begin{equation}\label{VQD_costfunc}
    L_{\mr{VQD}}^{(k)}(\bm{\theta}_k) \coloneqq \bra{\psi(\bm{\theta}_k)}\hat{H}\ket{\psi(\bm{\theta}_k)}
    +\sum_{i=0}^{k-1}\beta_i|\braket{\psi_i|\psi(\bm{\theta}_k)}|^2,
\end{equation}
where $\bm{\theta}_k$ are variational parameters to be optimized, $\ket{\psi(\bm{\theta}_k)}$ is an ansatz state such as Eq.~\eqref{eq: ansatz state}, and $\beta_i$ is an appropriate positive number to guarantee the orthogonality between the optimized ansatz state and the eigenstates up to the $(k-1)$th.
Let $\bm{\theta}_k^*$ be the optimal parameters. Then, as a result of the optimization, $L_{\mr{VQD}}^{(k)}(\bm{\theta}_k^*)$ and $\ket{\psi(\bm{\theta}_k^*)}$ can be regarded as good approximations of the $k$th excited-state energy and the $k$th excited state, respectively. 
By incrementing $k$ from $k=1$ to the desired energy level, one can obtain any eigenstate/energy of $\hat{H}$. 
We note that the procedure of the VQD can also be viewed as finding the ground state of the modified Hamiltonian $\hat{H}^{\prime} = \hat{H} + \sum_{i=0}^{k-1} \beta_i \ket{\psi_i}\bra{\psi_i}$ by using the original VQE algorithm; that is, the cost function is the expectation value of $\hat{H}^{\prime}$, $L_{\mr{VQD}}^{(k)}(\bm{\theta}_k) = \braket{\psi(\bm{\theta}_k)|\hat{H}^{\prime}|\psi(\bm{\theta}_k)}$.

\subsection{\label{subsec:constraints} Constrained VQE/VQD}
As explained in the introduction, the Hamiltonian to which we want to apply the VQE/VQD often has some symmetry, and 
it is essential to compute eigenvalues/states in the desired symmetry sector.
For such a purpose, we can introduce a penalty term to the cost function of the VQE/VQD algorithms~\cite{mcclean2016theory,Rubin_2018, greenediniz2019generalized,higgott2019variational,Ilya2019}, which we call the constrained VQE/VQD.
This can be interpreted as the penalty method studied in the field of computational optimization~\cite{Zak2013}.
Note that a regularization term added to prevent the overfitting of a cost function in machine learning can also be interpreted as a penalty term~\cite{shalev2014}. However, the regularization term in machine learning mostly constrains the functional complexity (\textit{i.e.,} the number of fitting parameters) of the cost function whereas the penalty term in the constrained VQE/VQD aims to restrict the symmetry of the wavefunction that the ansatz circuit produces.

Let $\hat{C}$ be an observable (a Hermitian operator) corresponding to the conserved quantity associated with the symmetry~\footnote{
The symmetry is expressed by a group of unitary operators that keep the system unchanged.
If the group is continuous (or a Lie group), $\hat{C}$ can be taken as one of the generators of the Lie group.
If the group is discrete, (\textit{e.g.}, the point-group symmetry), $\hat{C}$ can be taken as a $\hat{U}_g + \hat{U}_g^\dag$ and $i(\hat{U}_g - \hat{U}_g^\dag)$, where $\hat{U}_g$ is some generator(s) of the group that distinguishes the irreducible representation of the group.
}; $[\hat{H}, \hat{C}]=0$.
For example, when the system has $\textup{U}(1)$ symmetry, there is the particle-number conservation law, and $\hat{C}$ is the particle-number operator $\hat{N}$.
Suppose that we want to find an eigenstate of $\hat{H}$ that is also an eigenstate of $\hat{C}$ with the eigenvalue $c$. Here, $c$ determines the symmetry sector.
For the constrained VQE/VQD, two types of penalty terms and the cost function have been proposed:
\begin{align}
    \label{eq:cost_func1}
    &F^{(1)}(\bm{\theta}) \coloneqq L(\bm{\theta}) + \mu_C\bra{\psi(\bm{\theta})}(\hat{C} - c)^2\ket{\psi(\bm{\theta})},\\
    \label{eq:cost_func2}
    &F^{(2)}(\bm{\theta}) \coloneqq L(\bm{\theta}) + \mu_C(\bra{\psi(\bm{\theta})}\hat{C}\ket{\psi(\bm{\theta})} - c)^2, 
\end{align}
where $L(\bm{\theta})$ is either $L_\mr{VQE}(\bm{\theta})$ or $L_\mr{VQD}^{(k)}(\bm{\theta})$ for the constrained VQE/VQD, $\ket{\psi(\bm{\theta})}$ is an ansatz state with parameters such as Eq.~\eqref{eq: ansatz state}, and $\mu_C$ is some positive number to penalize the deviation of the expectation value of $\hat{C}$ for $\ket{\psi(\bm{\theta})}$ from the desired value $c$. 
(See Ref.~\cite{mcclean2016theory} for Eq.~\eqref{eq:cost_func1}; see Ref.~\cite{Ilya2019} for Eq.~\eqref{eq:cost_func2}.)

Intuitively, when $\mu_C$ is infinitely large, the second terms of the right hand sides of Eqs.~\eqref{eq:cost_func1} and \eqref{eq:cost_func2} get also infinitely large if $\bra{\psi(\bm{\theta})}\hat{C}\ket{\psi(\bm{\theta})} \neq c$.
The minimum of the cost function may be obtained at $\bm{\theta}^*$ satisfying $\bra{\psi(\bm{\theta}^*)}\hat{C}\ket{\psi(\bm{\theta}^*)} = c$ and the resulting state $\ket{\psi(\bm{\theta}^*)}$ may become an eigenstate of $\hat{C}$ with the eigenvalue $c$.
However, in the previous studies, there is no concrete mathematical discussion how the penalty terms in Eqs.~\eqref{eq:cost_func1} and \eqref{eq:cost_func2} work to guarantee that the minimum of the cost functions is reached at the state satisfying $\hat{C} \ket{\psi(\bm{\theta}^*)} = c\ket{\psi(\bm{\theta}^*)}$.
Moreover, it is known in the field of computational optimization that the large coefficient $\mu_C$ slows the convergence of the optimization process~\cite{Zak2013}.
The penalty coefficient $\mu_C$ is preferably small as long as one can ensure that the resulting state of the optimization is an eigenstate of $\hat{C}$ with the eigenvalue $c$.

\section{\label{sec:results} Main result}
We present our main results in this section.
We theoretically analyze the two cost functions~\eqref{eq:cost_func1} and \eqref{eq:cost_func2} for the constrained VQE/VQD, focusing on the role of the penalty terms and the effect of the values of $\mu_C$ on the optimization result.
For $F^{(1)}(\bm{\theta})$, we derive a rigorous condition of the penalty coefficient $\mu_C$ that guarantees that the result of the optimization yields a quantum state having the desired eigenvalues of $\hat{C}$.
Utilizing the condition, we also derive a convenient formula to set an appropriate value of $\mu_C$ in practical calculations.
For $F^{(2)}(\bm{\theta})$, we prove that this cost function does not work at all in a rigorous sense; a quantum state after the optimization always deviates slightly from the eigenstate of $\hat{C}$ for {\it any} finite value of $\mu_C$.
We show that the cost function $F^{(2)}(\bm{\theta})$ gives completely wrong results for some cases.
At the end of this section, we briefly discuss the effect of the noise in quantum circuits and measurements on those analyses of the cost functions.

Before presenting our main results, let us clarify the notation.
We consider a Hamiltonian $\hat{H}$ on an $n$-qubit Hilbert space and a conserved quantity $\hat{C}$ which commutes with the Hamiltonian, $[\hat{H},\hat{C}]=0$.
We write the spectral decomposition of $\hat{H}$ and $\hat{C}$ as 
\begin{equation}\label{eq:spec_decomp}
    \hat{H} \coloneqq \sum_{i=0}^{2^n-1} E_i\ket{i}\bra{i}, \quad
    \hat{C} \coloneqq \sum_{i=0}^{2^n-1} C_i\ket{i}\bra{i}, 
\end{equation}
where $\ket{i}$ is an eigenstate of $\hat{H}$ with eigenvalue $E_i$ ($E_0 \leqq E_1 \leqq \cdots \leqq E_{2^n-1}$) and $C_i$ is the eigenvalue of $\ket{i}$ for $\hat{C}$.
We want to obtain the ground-state (excited-state) energy of the sector of $C_i = c$ by the constrained VQE (VQD).
Since the VQD can be viewed as a search for the ground state of the modified Hamiltonian $\hat{H}^{\prime}$ as explained in Sec.~\ref{subsec:VQD}, here we assume that the target state is the ground state of the sector of $C_i = c$.
We denote the target state energy by $E_{i_0}$, which means that $C_i \neq c$ for $i=0,\ldots,i_0-1$. 

In addition, because our purpose is to explore the performance of the cost functions~\eqref{eq:cost_func1} and \eqref{eq:cost_func2} itself, we assume that the ansatz state $\ket{\psi(\bm{\theta})}$ (Eq.~\eqref{eq: ansatz state}) can represent an arbitrary $n$-qubit state.
That is, we expand the ansatz states as \begin{equation} \label{eq: expand ansatz}
 \ket{\psi(\bm{\theta})} = \sum_{i=0}^{2^n-1} a_i \ket{i},
\end{equation}
with $a_i \in \mathbb{C}$ and consider $a_i$ as taking any value satisfying $\sum_i |a_i|^2 = 1$.

\subsection{\label{subsec: construction_mu} Analysis of \texorpdfstring{$F^{(1)}$}{F1}: Formula for \texorpdfstring{$\mu_C$}{penalty coefficient}}
Here, we investigate how the penalty term in $F^{(1)}(\bm{\theta})$ works.
By substituting Eq.~\eqref{eq: expand ansatz} into the cost function~\eqref{eq:cost_func1}, we have
\begin{equation}\label{eq:costfunc_exp}
    F^{(1)}(\bm{\theta}) = \sum_{i=0}^{2^n-1} |a_i|^2(E_i + \mu_C (C_i - c)^2).
\end{equation}
To obtain $E_{i_0}$ as a result of the minimization of Eq.~\eqref{eq:costfunc_exp}
with respect to the parameters $\{ |a_i|^2 | \sum_i |a_i|^2 = 1 \}_{i=0}^{2^n-1}$,  it must hold that
\begin{equation}
    E_i + \mu_C(C_i - c)^2 \geqq E_{i_0}
\end{equation}
for all $i < i_0$ because we set $E_0 \leqq \ldots \leqq E_{i_0}, C_i \neq c \:(i=0,\cdots,i_0-1)$, and $\mu_C > 0$.
Therefore, it suffices to take
\begin{equation}\label{eq:mu_formula}
    \mu_C \geqq
    \max_{i<i_0} \frac{E_{i_0} - E_i}{(C_i - c)^2}. 
\end{equation}
In other words, if one takes $\mu_C$ satisfying Eq.~\eqref{eq:mu_formula}, it is guaranteed that the cost function~\eqref{eq:cost_func1} works properly to choose the desired quantum state that has the eigenvalue $c$ for $\hat{C}$ as a result of the minimization (with the parameters $|a_{i_0}|^2=1, |a_{i\neq i_0}|^2=0$).

Moreover, we can derive a formula for $\mu_C$ that is a little loose but much simpler than Eq.~\eqref{eq:mu_formula}. 
Let $C_{\min}$ be the smallest gap among distinct eigenvalues of $\hat{C}$.
Then, the right-hand side of \eqref{eq:mu_formula} can be upper-bounded as
\begin{equation}
    \max_{i<i_0}\frac{E_{i_0} - E_i}{(C_i - c)^2} 
    \leqq \max_{i<i_0}\frac{E_{i_0} - E_i}{C_{\min}^2}
    = \frac{E_{i_0} - E_0}{C_{\min}^2},
\end{equation}
and we obtain a convenient formula for $\mu_C$:
\begin{equation}\label{eq:estimate_mu}
 \mu_C \geqq \mu_C^{\mr{(simple)}} \coloneqq \frac{E_{i_0} - E_0}{C_{\min}^2}.
\end{equation}
This simple equation is one of our main results.
In the following, we explain how to estimate the right hand side of the equation in practical calculations.

The smallest gap $C_{\min}$ is uniquely determined for each conserved quantity $\hat{C}$, and it is easy to compute $C_{\min}$ for most of the cases we consider in quantum chemistry and condensed matter physics.
For example, when $\hat{C}$ is the particle number operator $\hat{N}$, we have $C_{\min} = 1$ because its eigenvalues are non-negative integers.
For the total spin-squared operator $\hat{S}^2$, since its eigenvalues are $S(S+1)$ for $S = 0,1/2,1,\ldots$, it follows that $C_{\min} = 3/4$. 
For the $z$-component of the total spin operator $\hat{S}_z$, we have $C_{\min} = 1/2$ because its eigenvalues are $0,\pm 1/2, \pm 1, \ldots$.
We use these values for numerical simulations in Sec.~\ref{sec:simulation}.
We note that because we consider the $n$-qubit Hilbert space, the eigenvalues of $\hat{C}$ are always discrete and $C_{\min} > 0$.

On the other hand, there are several ways to estimate the numerator of Eq.~\eqref{eq:estimate_mu}, or the energy gap between the desired state and the ground state, $E_{i_0} - E_0$. 
One way is to replace $E_{i_0}$ and $E_0$ with rough estimations calculated by classical computational methods that are less costly.
For example, in the case of quantum chemistry, we can use the Hartree-Fock (mean-field) method or the M{\o}ller–Plesset method~\cite{szabo2012modern}, which are much less costly on classical computers than the exact diagonalization method, to estimate $E_{i_0}$ and $E_0$.
We call the coefficient $\mu_C$ determined by a classical method $\mu_C^{(\mathrm{ce})}$:
\begin{equation}\label{eq:classical_estimate_mu}
    \mu_C^{(\mr{ce})} \coloneqq \frac{E_{i_0}^{(\mr{ce})} - E_0^{(\mr{ce})}}{C_{\min}^2},
\end{equation}
where $E_{i_0}^{(\mr{ce})}, E_0^{(\mr{ce})}$ are classically-estimated values.
Another way to estimate $E_{i_0} - E_0$ is to use the rigorous upper bound of it.
Let $\hat{H} = \sum_j c_jP_j$ be a decomposition of a given Hamiltonian into Pauli operators.
When we perform the VQE/VQD, this decomposition is already obtained~\cite{peruzzo2014variational, mcclean2016theory}.
By denoting the spectral norm of operators as $\|\cdot\|$, it follows that
$E_{i_0} - E_0 \leqq 2\|\hat{H}\| \leqq 2\sum_j |c_j|$ because $\|P_j\|=1$.
In this way we define another choice of $\mu_C$,
\begin{equation}\label{eq:loose_estimate_mu}
   \mu_C^{(\mathrm{rough})} \coloneqq \frac{2}{(C_{\min})^2}\sum_j |c_j|,
\end{equation}
which is easy and applicable to any system but may be too large for the fast convergence of the constrained VQE/VQD (see Sec.~\ref{sec:simulation}).

We note that the classical estimation $\mu_C^\mr{(ce)}$ may suffer from the deviation from the exact value of $\mu_C^{\mr{(simple)}}$.
When we overestimate the energy gap $E_{i_0} - E_0$ (and $\mu_C^{\mr{(simple)}}$), the cost function~\eqref{eq:cost_func1} with $\mu_C^{\mr{(ce)}}$ still works properly and we can find the desired eigenstate.
When we underestimate $E_{i_0} - E_0$, it is possible that the minimum of the cost function differs from the desired eigenstate.
We can nevertheless know whether such cases happen or not because we compute the value of $\braket{\psi(\bm{\theta})|(\hat{C}-c)^2|\psi(\bm{\theta})}$ during the optimization; when the value deviates from 0 drastically even after the optimization, we judge that the optimization fails to obtain the desired eigenstate.
We can then adopt slightly larger $\mu_C$ for the optimization.

Finally, we discuss several extensions of our result.
First, our result applies to a case where we have multiple conserved quantities $\{\hat{C}^{(l)}\}_l$.
The cost function in this case is
\begin{equation}
 \bra{\psi(\bm{\theta})}\hat{H}\ket{\psi(\bm{\theta})} + \sum_{l}\mu_{C^{(l)}}\bra{\psi(\bm{\theta})}(\hat{C}^{(l)} - c^{(l)})^2\ket{\psi(\bm{\theta})}. 
\end{equation}
The same discussion deriving Eq.~\eqref{eq:estimate_mu} leads to the condition for the optimized result to yield the eigenstate of $\hat{C}^{(l)}$ with eigenvalue $c^{(l)}$:
\begin{equation}
    \mu_{C^{(l)}} \geqq \frac{E_{i_0} - E_0}{(C^{(l)}_{\min})^2},
\end{equation}
where $C^{(l)}_{\min}$ is the smallest gap among distinct eigenvalues of $\hat{C}^{(l)}$.
Second, the formula~\eqref{eq:estimate_mu} is also applicable to the other VQE-based algorithms to obtain the excited states, namely, SSVQE~\cite{nakanishi2018subspace} and MCVQE~\cite{Parrish2019}.
The result is given by simply replacing $E_{i_0}$ of Eq.~\eqref{eq:estimate_mu} with the largest energy $E_{\mr{ex}}$ that one wants to obtain as a result of the optimization.
 
In summary, we derive the formulas~\eqref{eq:mu_formula} and \eqref{eq:estimate_mu} for $\mu_C$ in $F^{(1)}(\bm{\theta})$ that apply to general quantum systems, and they always guarantee that we can obtain the desired eigenstates/energies as a result of the optimization of $F^{(1)}(\bm{\theta})$.

\subsection{\label{subsec:theory_f_2}  Analysis of $F^{(2)}$: Failure to obtain the desired eigenstates}
In this section, we investigate the cost function $F^{(2)}(\bm{\theta})$ (Eq.~\eqref{eq:cost_func2}) and prove that we cannot obtain the exact desired energy/eigenstate by minimizing this cost function.
Concretely, the resulting energy after the minimization of $F^{(2)}(\bm{\theta})$ deviates from the one that we want to obtain by $O(1/\mu_C)$ even in the best cases and can be completely wrong in the worst cases.

Substituting Eq.~\eqref{eq: expand ansatz} into Eq.~\eqref{eq:cost_func2} leads to 
\begin{equation}\label{eq:exp_val_F2}
     F^{(2)}(\bm{\theta}) = \sum_{i = 0}^{2^n-1} |a_i|^2E_i + \mu_C\left(\sum_{i=0}^{2^n-1}|a_i|^2C_i - c\right)^2.
\end{equation}
We minimize this cost function with respect to the parameters $\{ |a_i|^2 | \sum_i |a_i|^2 = 1 \}_{i=0}^{2^n-1}$ and see whether $E_{i_0}$ is obtained as a result of the optimization (with the parameters $|a_{i_0}|^2=  1, |a_{i\neq i_0}|^2=0$).

\begin{figure}
    \centering
    \includegraphics[width = \linewidth]{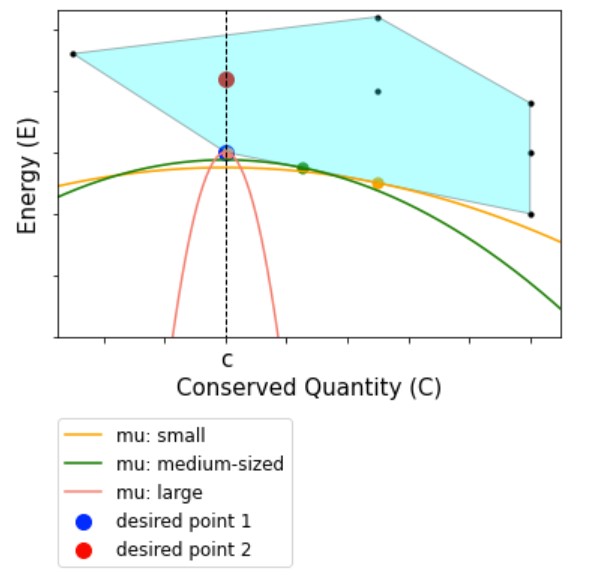}
    \caption{A schematic diagram of the location of the global minimum of the cost function $F^{(2)}(\bm{\theta})$. In this diagram, small black points correspond to simultaneous eigenstates for $\hat{H}$ and $\hat{C}$ (Eq.~\eqref{eq:spec_decomp}), and the convex envelope defined by them is colored cyan.
    The curves colored yellow, green, and pink are the parabolas~\eqref{eq:quadfunc_F2} with small, medium-sized, and large $\mu_C$, respectively.
    The tangent points to the convex envelope (\textit{i.e.}, the global minimum of the cost function) are also indicated by the points of the same color.  
    The dashed vertical line represents the desired expectation value of the conserved quantity.}
    \label{fig:F2_graph}
\end{figure}

To graphically see the way the magnitude of $\mu_C$ affects the global minimum of the cost function, we consider an orthogonal coordinate plane whose vertical axis represents the expectation values of $\hat{C}$ and whose horizontal axis represents the expectation values of $\hat{H}$, as depicted in Fig.~\ref{fig:F2_graph}.
For example, a point corresponding to some state $\ket{\phi}$ on this plane is $(\braket{\phi|\hat{C}|\phi}, \braket{\phi|\hat{H}|\phi})$.
Hereafter, we call this coordinate plane the $(C,E)$ plane.
For the ansatz state $\ket{\psi(\bm{\theta})}$ (Eq.~\eqref{eq: expand ansatz}), the expectation values are
\begin{align}
    \bra{\psi(\bm{\theta})} \hat{C} \ket{\psi(\bm{\theta})} = \sum_{i=0}^{2^n-1} |a_i|^2C_i, \\
    \bra{\psi(\bm{\theta})} \hat{H} \ket{\psi(\bm{\theta})} = \sum_{i=0}^{2^n-1} |a_i|^2E_i,
\end{align}
so all possible points corresponding to $\ket{\psi(\bm{\theta})}$ on the $(C, E)$ plane constitute the convex envelope defined by the $2^n$ points, $\{(C_0,E_0), (C_1,E_1),\ldots,(C_{2^n-1},E_{2^n-1})\}$ (the region colored cyan in Fig.~\ref{fig:F2_graph}). 
On this $(C,E)$ plane, a contour line of the cost function~\eqref{eq:exp_val_F2} (\textit{i.e.}, a set of points where Eq.~\eqref{eq:exp_val_F2} takes the same value) is a parabola:
\begin{equation}\label{eq:quadfunc_F2}
    E = -\mu_C (C - c)^2 + f,
\end{equation}
where $f$ is the value of Eq.~\eqref{eq:exp_val_F2}.
Therefore, the minimization of the cost function is identical to finding the smallest $f$ such that the parabola~\eqref{eq:quadfunc_F2}  and the convex envelope defined by $\{(C_i, E_i)\}_{i=0}^{2^n-1}$ have a non-empty intersection.
The smallest $f$ is achieved when the parabola \eqref{eq:quadfunc_F2} touches the polygon at just one point as shown in Fig.~\ref{fig:F2_graph}.

Depending on the location of the point $(c, E_{i_0})$ corresponding to the desired eigenstate, we can consider two cases:
\begin{enumerate}
    \item[(A)] The point $(c, E_{i_0})$ is a boundary point of the convex envelope (the blue point in Fig.~\ref{fig:F2_graph}),
    \item[(B)] The point $(c, E_{i_0})$ is an interior point of the convex envelope (the red point in Fig.~\ref{fig:F2_graph}).
\end{enumerate}

In case (A), unless $i_0 = 0$ (the desired state is the ground state of the Hamiltonian),
the global minimum of the cost function is reached at a point slightly different from $(c, E_{i_0})$ (the green and yellow points in Fig.~\ref{fig:F2_graph}).
Let us write an edge of the convex envelope that is tangent to the parabola~\eqref{eq:quadfunc_F2} as $E-E_{i_0} = \alpha (C - c)$ with some real number $\alpha$ on the $(C, E)$ plane.
It is straightforward to show that
\begin{align}
    \label{eq:tangent_point_f2}
    (C_t, E_t) &= \left( c - \frac{\alpha}{2\mu_C}, E_{i_0} - \frac{\alpha^2}{2\mu_C} \right), \\
    \label{eq:fmin_f2}
    f_{\min} &= E_{i_0} - \frac{\alpha^2}{4\mu_C},
\end{align}
where $(C_t, E_t)$ is a coordinate of the tangent point and $f_{\min}$ is the value of $f$ in Eq.~\eqref{eq:quadfunc_F2} for $(C,E) = (C_t, E_t)$, \textit{i.e.}, the minimal value of the cost function.
Those equations mean that the expectation values $(\bra{\psi(\bm{\theta})} \hat{C} \ket{\psi(\bm{\theta})}, \bra{\psi(\bm{\theta})} \hat{H} \ket{\psi(\bm{\theta})})$ at the global minimum of the cost function~\eqref{eq:cost_func2} {\it always} deviate from the target ones $(c, E_{i_0})$ by $O(1/\mu_C)$ for any finite $\mu_C$.
Only for infinitely large $\mu_C$, does the tangent point become $(c, E_{i_0})$, and we get the desired eigenstate.
On the other hand, in case (B), the desired eigenstate can never be obtained as a result of the optimization even for infinitely large $\mu_C$; the desired point $(c, E_{i_0})$ has no chance to be a tangent point to the parabola~\eqref{eq:quadfunc_F2}.

Before ending this section, we give another intuitive explanation for the reason why $F^{(1)}(\bm{\theta})$ works properly to choose the desired eigenstate while $F^{(2)}(\bm{\theta})$ does not.
In the expression of $F^{(1)}(\bm{\theta})$ (the right hand side of Eq.~\eqref{eq:costfunc_exp}), both the ``energy part" (the first term) and the ``conserved-quantity" part (the second term) are proportional to $|a_i|^2$.
On the other hand, in $F^{(2)}(\bm{\theta})$ (the right hand side of Eq.~\eqref{eq:exp_val_F2}), the conserved-quantity part is a quadratic function of $|a_i|^2$ while the energy part is proportional to $|a_i|^2$. 
Since $|a_i|^2 \leqq 1$ for all $i$, the deviation in the conserved-quantity part from the desired value is less penalized for $F^{(2)}(\bm{\theta})$ than for $F^{(1)}(\bm{\theta})$.
This makes the difference between the performance of the cost functions $F^{(1)}(\bm{\theta})$ and $F^{(2)}(\bm{\theta})$.
Indeed, the minimum of $F^{(2)}(\bm{\theta})$ in case (A) is achieved at a point where the value of the conserved quantity deviates from the desired one, and the energy gets smaller than the desired energy. 

In short, we show that the cost function $F^{(2)}(\bm{\theta})$ for the constrained VQE/VQD (Eq.~\eqref{eq:cost_func2}) does not work at all in a rigorous sense.
We can only obtain a slightly-deviated desired eigenstate with an error of $O(1/\mu_C)$ even in the best cases while we obtain a totally different eigenstate in the worst case.

\subsection{Analysis of noise robustness of $F^{(1)}$ and $F^{(2)}$}
Here, we investigate the effects of noise on the performance of the two cost functions of $F^{(1)}(\bm{\theta})$ and $F^{(2)}(\bm{\theta})$. 
To make analysis simple and obtain the general tendency of the robustness of $F^{(1)}(\bm{\theta})$ and $F^{(2)}(\bm{\theta})$ to noise, we consider the $n$-qubit depolarizing channel to represent the noise. 
The $n$-qubit depolarizing channel outputs the completely mixed state with a certain probability and outputs the original input state otherwise~\cite{nielsen_chuang_2010}.  
More formally, the $n$-qubit depolarizing channel $\Delta_p$ with probability $p$ is defined as
\begin{equation}
    \Delta_p(\rho) \coloneqq (1-p)\rho + p\frac{I}{2^n}
\end{equation}
for all $n$-qubit states $\rho$, where $0\leqq p \leqq 1$ and $I$ is the $2^n\times 2^n$ identity matrix. For the analysis of noise, we suppose that the depolarizing channel is applied to the ansatz state $\ket{\psi(\bm{\theta})}$ resulting in the mixed state
\begin{equation}~\label{eq:noisy_ansatz_state}
    \begin{aligned}
    \rho_p(\bm{\theta}) 
    &\coloneqq \Delta_p(\ket{\psi(\bm{\theta})}\bra{\psi(\bm{\theta})})\\
    &=(1-p)\ket{\psi(\bm{\theta})}\bra{\psi(\bm{\theta})} + p\frac{I}{2^n}. 
    \end{aligned}
\end{equation}
When $p=1$, the resulting state is the completely mixed state $I/2^n$, which does not depend on the parameters. In this case, the cost function is a constant function, and the optimization becomes trivial. Hereafter, we take $0\leqq p < 1$ to avoid this situation.  

First, we investigate the effects of this noise on $F^{(1)}(\bm{\theta})$. 
Let $\tilde{F}^{(1)}(\bm{\theta})$ be the modified version of $F^{(1)}(\bm{\theta})$ where the expectation values with respect to $\ket{\psi(\bm{\theta})}$ are replaced with those with respect to $\rho_p(\bm{\theta})$.
Then, considering the expansion of $\ket{\psi(\bm{\theta})}$ as in Eq.~\eqref{eq: expand ansatz} and the noisy ansatz state~\eqref{eq:noisy_ansatz_state}, we obtain 
\begin{equation}
    \begin{aligned}
    \tilde{F}^{(1)}(\bm{\theta}) 
    &= (1-p)\sum_{i=0}^{2^n-1} |a_i|^2(E_i + \mu_C (C_i - c)^2) \\
    &+ p\tr(\hat{H}+\mu_C(\hat{C}-c)^2).
    \end{aligned}
\end{equation}
Since the second term of $\tilde{F}^{(1)}(\bm{\theta})$ does not depend on the parameters $\bm{\theta}$, the optimization of $\tilde{F}^{(1)}(\bm{\theta})$ with respect to $\bm{\theta}$ can be completed by the optimization of the first term. The first term is identical to $F^{(1)}(\bm{\theta})$ up to the constant coefficient $(1-p)$; therefore, the optimization of $\tilde{F}^{(1)}(\bm{\theta})$ yields the same optimal parameters as that of $F^{(1)}(\bm{\theta})$.
Thus, in this case, the noise has no effect and we can obtain the desired eigenstate if we set the penalty coefficient $\mu_C$ appropriately as in Eq.~\eqref{eq:mu_formula}. 

Next, we analyze the effects of the $n$-qubit depolarizing noise on $F^{(2)}(\bm{\theta})$.
Similarly, consider the modified version $\tilde{F}^{(2)}(\bm{\theta})$ where the expectation values are computed with respect to $\rho_p(\bm{\theta})$.
By using the expansions of $\ket{\psi(\bm{\theta})}$ (Eq.~\eqref{eq: expand ansatz}) and the noisy ansatz state (Eq.~\eqref{eq:noisy_ansatz_state}), we obtain 
\begin{equation}\label{eq:exp_val_F2_noise}
    \begin{aligned}
    \tilde{F}^{(2)}(\bm{\theta}) 
    &= (1-p)\sum_{i = 0}^{2^n-1} |a_i|^2E_i + p\tr(\hat{H}) \\
    &+ \mu_C\left((1-p)\sum_{i=0}^{2^n-1}|a_i|^2C_i - (c - p\tr(\hat{C}))\right)^2.
    \end{aligned}
\end{equation}
A contour line of the cost function $\tilde{F}^{(2)}(\bm{\theta})$ on the $(C,E)$ plane considered in Sec.~\ref{subsec:comparison_F1F2} becomes a parabola again:
\begin{equation}\label{eq:quadfunc_F2_noise}
    \begin{aligned}
    (1-p)E = 
    &-\mu_C ((1-p)C - (c - p\tr(\hat{C})))^2 \\
    &+ \tilde{f} - p\tr(\hat{H}),
    \end{aligned}
\end{equation}
where $\tilde{f}$ is the value of Eq.~\eqref{eq:exp_val_F2_noise}.
The minimum of $\tilde{F}^{(2)}(\bm{\theta})$ is achieved when the parabola \eqref{eq:quadfunc_F2_noise} touches the convex envelope constructed from the set of points $\{(C_i,E_i)\}_i$ corresponding to the eigenvalues of $\hat{C}$ and $\hat{E}$. 
Intuitively, observing that the axis of symmetry of the parabola deviates from $C=c$, we expect that the optimization of $\tilde{F}^{(2)}(\bm{\theta})$ does not yield the desired eigenstate whose eigenvalue for $\hat{C}$ is $c$.
Indeed, it is straightforward to obtain 
\begin{align*}
    \tilde{C}_t &= \frac{1}{1-p}\left((c-p\tr(\hat{C})) - \frac{\alpha}{2\mu_C}\right) \\
    &= C_t + p\left(c-\tr(\hat{C})-\frac{\alpha}{2\mu_C}\right) + O(p^2),\\
    \tilde{E}_t &= E_{i_0} - \frac{\alpha}{1-p}\left(\frac{\alpha}{2\mu_C} + p\tr(\hat{C})\right)\\
    &= E_t -p\left(\alpha\tr(\hat{C})+\frac{\alpha^2}{2\mu_C}\right) + O(p^2)\\
    \tilde{f}_{\min} &= f_{\min} + p(\tr(\hat{H})-\alpha\tr(\hat{C})-E_{i_0}),
\end{align*}
where $(\tilde{C}_t, \tilde{E}_t)$ is a coordinate of the tangent point and $\tilde{f}_{\min}$ is the value of $\tilde{F}^{(2)}(\bm{\theta})$ at that point.
We again write an edge of the convex envelope that is tangent to the parabola as $E-E_{i_0} = \alpha (C - c)$. 
In addition, $(C_t,E_t)$ and $f_{\min}$ are the values in the noiseless case as shown in Eqs.~\eqref{eq:tangent_point_f2} and \eqref{eq:fmin_f2}.
The above equations tell us that the optimal parameters that minimize $\tilde{F}^{(2)}(\bm{\theta})$ deviate from those minimizing $F^{(2)}(\bm{\theta})$ due to the non-zero probability $p$ of the depolarizing noise.
This is in stark contrast to the case of $F^{(1)}(\bm{\theta})$ and $\tilde{F}^{(1)}(\bm{\theta})$.

We have shown that $F^{(1)}(\bm{\theta})$ is robust to the $n$-qubit depolarizing while $F^{(2)}(\bm{\theta})$ is not. 
We expect that this tendency of the noise vulnerability of the two cost functions applies to more general types of the noise, although the further detailed analysis is needed.

\section{\label{sec:simulation} Numerical simulations}
In this section, we numerically simulate the constrained VQE/VQD for two molecules, $\ce{H_2}$ and $\ce{H_4}$, to validate our results presented in the previous section.
We also discuss the comparison of the performance of the two cost functions $F^{(1)}(\bm{\theta})$ and $F^{(1)}(\bm{\theta})$ in practical calculations. 

Our setup for numerical simulations is as follows.
We consider a hydrogen molecule $\ce{H_2}$ at bond distance $R=\SI{0.7414}{\AA}$ and a hydrogen chain $\ce{H_4}$ where four hydrogen atoms are aligned in line with the identical bond distance $R=\SI{2.0}{\AA}$.
We adopt the STO-3G minimal basis set to perform the restricted Hartree-Fock calculation for the two molecules.
We prepare the fermionic second-quantized Hamiltonian for electrons using PySCF~\cite{Sun2018_pyscf} and Openfermion~\cite{mcclean2017openfermion} and use Jordan-Wigner transformation~\cite{Jordan1928} to map the fermionic Hamiltonians into qubit Hamiltonians~\cite{mcardle2018quantum, Cao2018}.
The numbers of qubits to express the Hamiltonian are $4$ and $8$ for $\ce{H_2}$ and $\ce{H_4}$, respectively.
As conserved quantities, we consider the total number of electrons $\hat{N}_e$, the total spin-squared operator $\hat{S}^2$, and the total $z$-component of the spin $\hat{S}_z$.
Those conserved quantities are also transformed into operators on qubits by Jordan-Wigner transformation.

\begin{figure}
    \centering
    \includegraphics[width = 0.8\linewidth]{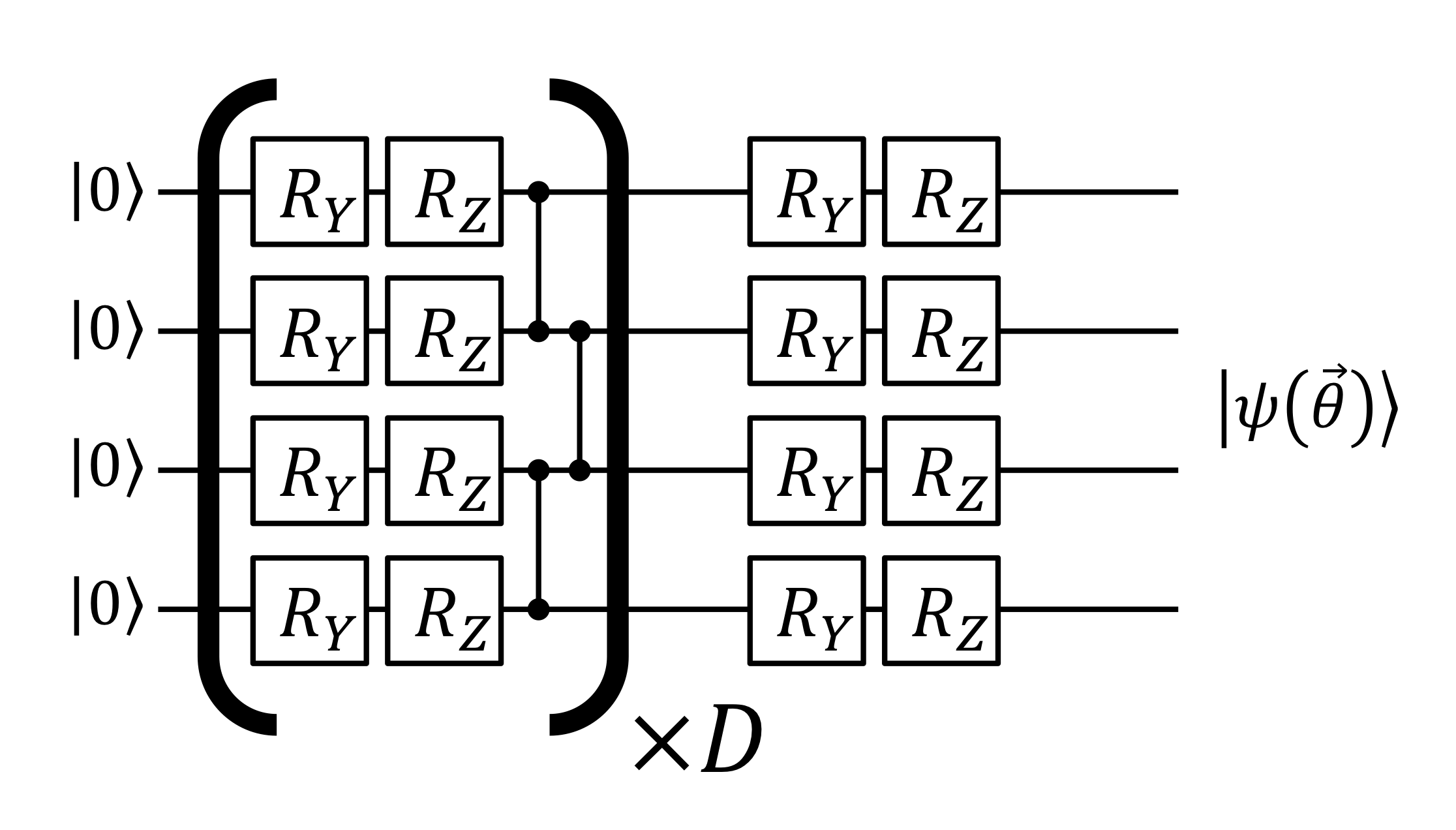}
    \caption{Quantum circuit for the hardware-efficient type ansatz.
    Each of $R_Y = e^{i\theta Y/2}$ and $R_Z = e^{i\theta Z/2}$ gates has an independent parameter $\theta$, where $Y$ and $Z$ are the Pauli $Y, Z$ operators.
    $D$ denotes the depth of the ansatz.}
    \label{fig:HEA}
\end{figure}

We employ the hardware-efficient type ansatz~\cite{kandala2017hardware} shown in Fig.~\ref{fig:HEA} for $\ce{H_2}$. 
The depth of the ansatz is $D=4$ in the constrained VQE to compute the $\ce{T_1}$ state (defined later) and $D=12$ in the constrained VQD to compute the $\ce{S_1}$ state (defined later). 
We adopt the real-valued symmetry-preserving type ansatz of the depth $D=12$ introduced in Refs.~\cite{ibe2020calculating,Gard2019} for $\ce{H_4}$ to perform the constrained VQE/VQD.
We do not include any noise for quantum circuit simulations, and exact expectation values are used in numerical simulations.
All simulations are performed by the high-speed quantum circuit simulator Qulacs~\cite{qulacs_2018}.

\subsection{\label{subsec:comparison_mu} Numerical demonstration for analysis on $F^{(1)}$}
We perform simulations to validate the analysis on $F^{(1)}(\bm{\theta})$ in Sec.~\ref{subsec: construction_mu} and investigate how the magnitude of penalty coefficient $\mu_C$ affects the accuracy and the convergence of the optimization.

As for the target state of the constrained VQE/VQD, we consider the $\ce{T_1}$ state (the ground state in the triplet sector) which corresponds to $\hat{S}^2 = 2$ and $\hat{S}_z = -1$ and the $\ce{S_1}$ state (the first excited state in the singlet sector) which corresponds to $\hat{S}^2 = 0$ and $\hat{S}_z = 0$. 
Note that the ground state is the spin singlet ($\hat{S}^2 = 0$ and $\hat{S}_z = 0$) state for both the $\ce{H_2}$ and $\ce{H_4}$ molecules.
We set the penalty terms for some of $\hat{N}, \hat{S}^2$ and, $\hat{S}_z$ depending on the molecule and the target state, which is summarized in TABLE~\ref{table_penalty_summary}.
We compute the energy of the $\ce{S_1} (\ce{T_1})$ state as the ground state (first excited-state) energy of the constrained VQE (VQD) with these constraints. 

\begin{table}
    \centering
    \begin{tabular}{c c c c c} 
    \hline 
     && $\ce{H_2}$ && $\ce{H_4}$ \\ \hline
     triplet $\ce{T_1}$ && $\hat{N}_e=2, \hat{S}^2=2, \hat{S}_z=-1$ &&  $\hat{S}^2=2, \hat{S}_z=-1$ \\
     singlet $\ce{S_1}$ && $\hat{N}_e=2, \hat{S}^2=0$ &&  $\hat{S}^2=0$ \\ \hline 
    \end{tabular}
    \caption{The penalty terms included in the cost functions of the constrained VQE/VQD in numerical simulations. Since the real-valued symmetry-preserving type ansatz~\cite{ibe2020calculating, Gard2019} preserves the number of electrons of the reference state (see Eq.~\eqref{eq: ansatz state}) and we take $\ket{\psi_0} = \ket{00001111}$, we do not set the penalty on $\hat{N}_e$ for $\ce{H_4}$.  
    \label{table_penalty_summary}}
\end{table}

First, we estimate $\mu_C^{(\mathrm{ce})}$ (Eq.~\eqref{eq:classical_estimate_mu}) and $\mu_C^{(\mathrm{rough})}$ (Eq.~\eqref{eq:loose_estimate_mu}) for the $\ce{H_2}$ and $\ce{H_4}$ molecules.
We use the configuration interaction singles and doubles (CISD) method implemented in PySCF~\cite{Sun2018_pyscf} to compute the classical estimations of the energies appearing in $\mu_C^{(\mathrm{ce})}$. 
Note that we may use other less-costly methods for large molecules. 
The values of $\mu_C$ are presented in TABLE~\ref{table_mu}.
As expected, $\mu_C^{(\mathrm{rough})}$ is much larger than $\mu_C^{(\mathrm{ce})}$ for all coefficients, which may cause the slow convergence of the optimization in the constrained VQE/VQD.

Next, by using the values of $\mu_C$ in TABLE~\ref{table_mu}, we numerically simulate the constrained VQE/VQD with the cost function $F^{(1)}(\bm{\theta})$ (Eq.~\eqref{eq:cost_func1}).
The optimization is performed by several optimizers (Powell, conjugate gradient (CG), and Broyden-Fletcher-Goldfarb-Shanno (BFGS)) implemented in SciPy, a numerical library of Python~\cite{Virtanen_2020} with default parameters.
We simulate the constrained VQE (VQD) to obtain the $\ce{T_1}$ state ($\ce{S_1}$ state) for both the $\ce{H_2}$ and $\ce{H_4}$ molecules.
When performing the constrained VQD for the $\ce{S_1}$ state, we take a hyperparameter $\beta_0$ (see Eq.~\eqref{VQD_costfunc}) as $\beta_0 = 2(E_{\ce{S_1}} - E_0)$, where $E_{\ce{S_1}}$ and $E_0$ are energies of the $\ce{S_1}$ state and the ground state, respectively.
This choice of $\beta_0$ is based on the original VQD proposal~\cite{higgott2019variational} to ensure that the VQD cost function $L_{\mr{VQD}}(\bm{\theta})$ brings out the excited state properly.
In practice we also estimate the value of $\beta_0$ in the same way as we do for $\mu_C$, 
$ \beta_0^{(\mr{ce})} = 2(E_{\ce{S_1}}^{(\mr{ce})} - E_0^{(\mr{ce})})$ and $\beta_0^{(\mr{rough)}} = 4\sum_i |c_i|$, 
where the superscript ``(ce)" represents the classical estimation of the energy and $\{c_i\}_i$ are the coefficients appearing in the decomposition of the Hamiltonian into Pauli operators, $\hat{H} = \sum_i c_iP_i$. 
We adopt $\beta_0^{(\mr{ce})}$ for the simulation with $\mu_C^{(\mr{ce})}$ and $\beta_0^{(\mr{rough})}$ for that with $\mu_C^{(\mr{rough})}$.
We run the simulations for ten different initial values of $\bm{\theta}$ in each case and compute the average number of the evaluations of $F^{(1)}(\theta)$ and the average value of the expectation value of the Hamiltonian for the optimized ansatz state.

The results for $\ce{H_2}$ and $\ce{H_4}$ are shown in TABLEs~\ref{table_H2_mu} and \ref{table_H4_mu}, respectively.
Those results clearly indicate that the penalty coefficient $\mu_C$ determined by our formulas in Sec.~\ref{subsec: construction_mu} is sufficient for the cost function to choose the desired eigenstates.
Moreover, we observe that the number of function evaluations is much smaller for $\mu_C^{(\mathrm{ce})}$ than for $\mu_C^{(\mathrm{rough})}$.
The formula for $\mu_C^{(\mathrm{rough})}$ is convenient and always applicable as long as the Hamiltonian is known, but it may be inappropriate for the fast convergence in practical calculations.
We note that the optimization result is slightly worse for the Powell method possibly due to the imperfectness of the optimization, nevertheless the tendency of the small number of function evaluations in $\mu_C^{(\mathrm{ce})}$ is well observed.

\begin{table}
    \centering
    \begin{tabular}{c c c c c c c c} 
        \hline 
        &\multicolumn{4}{c}{$\mu_C^{(\mathrm{ce})}$} &&\multicolumn{2}{c}{$\mu_C^{(\mathrm{rough})}$}\\\hline
        &\multicolumn{2}{c}{$\ce{H_2}$} 
        &\multicolumn{2}{c}{$\ce{H_4}$} 
        &&$\ce{H_2}$ 
        &$\ce{H_4}$ \\\hline
        &Triplet T$_1$& Singlet S$_1$& Triplet &Singlet&\\\hline
    $\mu_{N}$ &$0.6048$ &$0.9674$ &$0.03421$ &$0.03711$ &&$3.968$ &$12.11$\\
    $\mu_{S^2}$ &$1.075$ &$1.720$ &$0.06082$ &$0.06597$ && $7.054$ &$21.52$\\
    $\mu_{S_z}$ &$2.419$ &$3.869$ &$0.1368$ &$0.1484$ &&$15.87$ &$48.42$\\
    \hline 
    \end{tabular}
    \caption{Penalty coefficients calculated by our formulas~\eqref{eq:classical_estimate_mu} and \eqref{eq:loose_estimate_mu} for the target eigenstates of $\ce{H_2}$ and $\ce{H_4}$.}
    \label{table_mu}
\end{table}
\begin{table*}
    \centering
    \begin{tabular}{c c r r r r c r r r r } 
    \hline
    && \multicolumn{4}{c}{$\ce{H_2}$ Triplet $\ce{T_1}$} && \multicolumn{4}{c}{$\ce{H_2}$ Singlet $\ce{S_1}$} \\\hline
    && \multicolumn{2}{c}{$\mu_C^{(\mathrm{ce})}$} & \multicolumn{2}{c}{$\mu_C^{(\mathrm{rough})}$}  && \multicolumn{2}{c}{$\mu_C^{(\mathrm{ce})}$} & \multicolumn{2}{c}{$\mu_C^{(\mathrm{rough})}$} \\\hline
    Optimizer && \multicolumn{1}{c}{nfev} &  \multicolumn{1}{c}{residuals} & \multicolumn{1}{c}{nfev} &  \multicolumn{1}{c}{residuals} && \multicolumn{1}{c}{nfev} & \multicolumn{1}{c}{residuals} & \multicolumn{1}{c}{nfev} &  \multicolumn{1}{c}{residuals}\\\hline
    Powell &&$\SI{3569}{}$ & $\SI{0.000011}{}$ & $\SI{5073}{}$ & $\SI{0.000002}{}$ && $\SI{35484}{}$ & $\SI{-0.000219}{}$ & $\SI{48863}{}$ & $\SI{0.064635}{}$ \\\hline
    CG && $\SI{5055}{}$ &$\SI{0.000000}{}$ & $\SI{7987}{}$ & $\SI{0.000000}{}$ && $\SI{142559}{}$ &$\SI{0.000000}{}$ & $\SI{263750}{}$ &$\SI{0.000000}{}$  \\\hline
    BFGS && $\SI{1406}{}$ &$\SI{0.000000}{}$ & $\SI{1615}{}$ &$\SI{0.000000}{}$ && $\SI{16128}{}$ &  $\SI{0.000000}{}$ & $\SI{19373}{}$ & $\SI{0.000000}{}$ \\\hline
    \end{tabular}
    \caption{The results of the numerical simulations for calculating the $\ce{T_1}$ state and the $\ce{S_1}$ state of $\ce{H_2}$ by the constrained VQE/VQD with the cost function $F^{(1)}(\bm{\theta})$ whose penalty coefficient is $\mu^{(\mathrm{ce})}$ or $\mu^{(\mathrm{rough})}$.
    The optimizer is chosen from the Powell, CG, and BFGS methods. 
    The term ``nfev" is the number of evaluations of the cost function during the optimization, and ``residuals" represents the difference between the expectation value of the Hamiltonian for the resulting optimized state and the theoretical (or desired) energy in Hartree (Ha).
    The theoretical value for the $\ce{T_1}$ state  is $\SI{-0.532479}{Ha}$, and that of the $\ce{S_1}$ state is $\SI{-0.169901}{Ha}$.
    We perform the constrained VQE/VQD for ten different initial values, and all values shown in this table are the average for the ten trials. 
    }
    \label{table_H2_mu}
\end{table*}
\begin{table*}
    \centering
    \begin{tabular}{c c r r r r c r r r r } 
    \hline
    && \multicolumn{4}{c}{$\ce{H_4}$ Triplet $\ce{T_1}$} && \multicolumn{4}{c}{$\ce{H_4}$ Singlet $\ce{S_1}$} \\\hline
    && \multicolumn{2}{c}{$\mu_C^{(\mathrm{ce})}$} & \multicolumn{2}{c}{$\mu_C^{(\mathrm{rough})}$} && \multicolumn{2}{c}{$\mu_C^{(\mathrm{ce})}$} & \multicolumn{2}{c}{$\mu_C^{(\mathrm{rough})}$} \\\hline
    Optimizer && \multicolumn{1}{c}{nfev} & \multicolumn{1}{c}{residuals} & \multicolumn{1}{c}{nfev}  & \multicolumn{1}{c}{residuals} && \multicolumn{1}{c}{nfev}  & \multicolumn{1}{c}{residuals} & \multicolumn{1}{c}{nfev} & \multicolumn{1}{c}{residuals}\\\hline
    Powell &&$\SI{20405}{}$ &$\SI{0.007091}{}$ & $\SI{110822}{}$ &$\SI{0.082902}{}$ && $\SI{42933}{}$  &$\SI{0.024028}{}$ & $\SI{266387}{}$ & $\SI{0.138162}{}$ \\\hline
    CG && $\SI{60563}{}$ &$\SI{0.000000}{}$ & $\SI{263041}{}$  &$\SI{0.009657}{}$ && $\SI{115915}{}$ & $\SI{0.000000}{}$ & $\SI{528539}{}$  &$\SI{0.020683}{}$ \\\hline
    BFGS && $\SI{17604}{}$  &$\SI{0.000000}{}$ & $\SI{147063}{}$ &$\SI{0.001421}{}$ && $\SI{31042}{}$  &$\SI{0.000000}{}$ & $\SI{252330}{}$ & $\SI{0.000000}{}$  \\\hline
    \end{tabular}
    \caption{
    The results of the numerical simulations for calculating the $\ce{T_1}$ state and the $\ce{S_1}$ state of $\ce{H_4}$ by the constrained VQE/VQD with the cost function $F^{(1)}(\bm{\theta})$ whose penalty coefficient is $\mu^{(\mathrm{ce})}$ or $\mu^{(\mathrm{rough})}$.
    The optimizer is chosen from the Powell, CG, and BFGS methods. 
    The term ``nfev" is the number of evaluations of the cost function during the optimization, and ``residuals" represents the difference between the expectation value of the Hamiltonian for the resulting optimized state and the theoretical (or desired) energy in Hartree (Ha).
    The theoretical value for the $\ce{T_1}$ state  is $\SI{-1.881876}{Ha}$, and that of the $\ce{S_1}$ state is $\SI{-1.856584}{Ha}$.
    We perform the constrained VQE/VQD for ten different initial values, and all values shown in this table are the average for the ten trials. }
    \label{table_H4_mu}
\end{table*}

\subsection{\label{subsec:comparison_F1F2} Numerical demonstration for analysis on $F^{(2)}$ and comparison of practical performance of two cost functions}
We perform the following two simulations to validate our analysis on $F^{(2)}(\bm{\theta})$ in Sec.~\ref{subsec:theory_f_2} and to compare the two cost functions $F^{(1)}(\bm{\theta})$ and $F^{(2)}(\bm{\theta})$.
\begin{itemize}
\item[(i)] Computing the energy of the $\ce{T_1}$ state for the $\ce{H_4}$ by the constrained VQE with the two cost functions having a range of penalty coefficients, $\mu_C = \mu_{S^2} = \mu_{S_z} = 0.01, 0.1, 1, 10, 100$. 
\item[(ii)] Computing the energy of the $\ce{S_1}$ state of $\ce{H_2}$ using the constrained VQD under the constraint $\hat{N}_e = 2$ with $\mu_C = \mu_{N_e} = 0.01, 0.1, 1, 10, 100$. 
Since we set the constraint only for $\hat{N}_e$ in this case, the $\ce{S_1}$ state is obtained as the fourth excited state of the constrained VQD $(k=4)$.
The hyperparameters of the VQD $\{\beta_i\}_{i=0}^3$ (see Eq.~\eqref{VQD_costfunc}) are all set to $3.0$.
\end{itemize}
For both simulations, we use the BFGS optimizer, which shows the best performance in Sec.~\ref{subsec:comparison_mu}. 

The results of simulation (i) are shown in TABLE~\ref{table_cost}.
We perform the constrained VQE for ten different initial values and calculate the average of the expectation values of the Hamiltonian for the resulting optimized states.
Unlike in the previous section, we calculate the average number of ``Pauli measurements" to assess the computational cost for the constrained VQE.
The number of Pauli measurements is defined as (the number of evaluations of the cost function during the optimization) $\times$ (the number of Pauli operators to be measured to evaluate the cost function once).
This number represents the actual computational cost to perform experiments on a real NISQ device~\cite{mcclean2016theory}. 

From TABLE~\ref{table_cost}, we can observe two facts.
First, the accuracy (``residuals") of the optimization tends to improve with $1/\mu_C$, which validates the discussion in Sec.~\ref{subsec:theory_f_2}.
Second, more importantly from the viewpoint of practical applications, the number of Pauli measurements for $F^{(2)}(\bm{\theta})$ is smaller than that for $F^{(1)}(\bm{\theta})$.
This is mainly because the evaluation of  $F^{(1)}(\bm{\theta})$ involves the evaluation of the expectation value of the operator $(\hat{C}-c)^2$ that consists of more Pauli operators than the original $\hat{C}$ does.
The evaluation of $F^{(2)}(\bm{\theta})$ necessitates only the expectation value of $\hat{C}$ itself.
If we choose the best cases for $F^{(1)}(\bm{\theta}) \: (\mu_C =0.01)$ and $F^{(2)}(\bm{\theta}) \: (\mu_C =100)$, where both cost functions give sufficiently accurate results, the number of Pauli measurements still is smaller for  $F^{(2)}(\bm{\theta})$.
This interesting observation indicates that even though $F^{(2)}(\bm{\theta})$ cannot theoretically achieve an exact desired energy, it may achieve a sufficiently accurate energy with small computational costs for some cases. 

Nevertheless, we also find a practical case where $F^{(2)}(\bm{\theta})$ can never achieve a target energy even with infinitely large $\mu_C$ in simulation (ii).
The results of simulation (ii) are shown in TABLE~\ref{table_F2_fail}.
The optimization of $F^{(1)}(\bm{\theta})$ gives the correct values for sufficiently large $\mu_C$. On the other hand, the optimization of $F^{(2)}(\bm{\theta})$ yields results crucially far from the exact value even for large $\mu_C$.
In this case, the target eigenstate is inside the convex envelope explained in Sec.~\ref{subsec:theory_f_2} and we simply cannot obtain that state by optimizing the cost function $F^{(2)}(\bm{\theta})$.

To summarize, we numerically validate our theoretical analysis on $F^{(2)}(\bm{\theta})$ in Sec.~\ref{subsec:theory_f_2} in practical quantum chemistry calculations.
We find examples for both cases (A) and (B) in Sec.~\ref{subsec:theory_f_2}, where one can obtain the desired energy with an error of $O(1/\mu_C)$ (case (A)) or cannot obtain it even with infinitely large $\mu_C$ (case (B)).
Moreover, we find that sometimes the cost function $F^{(2)}(\bm{\theta})$ can find a sufficiently accurate energy with less computational cost than $F^{(1)}(\bm{\theta})$, reflecting the difference in the amount of effort to evaluate the cost functions on NISQ devices.
We stress that the use of $F^{(2)}(\bm{\theta})$ may be preferable in some cases to reduce the computational cost, but whether we can obtain the target state as a result of the optimization cannot be guaranteed at all {\it a priori}.  

\begin{table}
    \centering
    \begin{tabular}{c c r r c r r} 
    \hline
    && \multicolumn{5}{c}{$\ce{H_4}$ Triplet $\ce{T_1}$}\\ \hline
    &&\multicolumn{2}{c}{$F^{(1)}(\bm{\theta})$} &&\multicolumn{2}{c}{$F^{(2)}(\bm{\theta})$} \\\hline
    $\mu_C$ && \multicolumn{1}{c}{$N_\mr{meas}$}  & \multicolumn{1}{c}{residuals} && \multicolumn{1}{c}{$N_\mr{meas}$}  & \multicolumn{1}{c}{residuals} \\\hline
    $0.01$ && $\SI{12772015}{}$ &$\SI{0.000000}{}$&&$\SI{4446580}{}$ & $\SI{-0.002530}{}$ \\\hline
    $0.1$ && $\SI{13315505}{}$  &$\SI{0.000000}{}$&&$\SI{4448356}{}$ &$\SI{-0.000253}{}$ \\\hline
    $1$ && $\SI{26973645}{}$  &$\SI{0.000000}{}$&&$\SI{4583370}{}$ & $\SI{-0.000025}{}$ \\\hline
    $10$ && $\SI{85835975}{}$  &$\SI{0.000044}{}$&&$\SI{6571274}{}$  &$\SI{-0.000002}{}$\\\hline
    $100$ && $\SI{118039912}{}$  &$\SI{0.015919}{}$&&$\SI{9781409}{}$ &$\SI{0.000000}{}$ \\\hline 
    \end{tabular}
    \caption{The results of the numerical simulations for calculating the $\ce{T_1}$ state of the $\ce{H_4}$ by the constrained VQE with various $\mu_C$.
    $N_\mr{meas}$ represents the number of Pauli measurements necessary to perform the whole optimization process, which is computed as (the number of evaluations of the cost function during the optimization) $\times$ (the number of Pauli operators to be measured to get the value of the cost function once).
    The term ``residuals" represents the difference between the expectation value of the Hamiltonian for the resulting optimized state and that for the exact (desired) state in Hartree (Ha).
    The exact value for this case is $\SI{-1.881876}{Ha}$.
    We perform the constrained VQE for ten different initial values, and all values shown in this table are the average for the ten trials. }
    \label{table_cost}
\end{table}

\begin{table}
    \centering
    \begin{tabular}{c c r r c r r} 
    \hline
    && \multicolumn{5}{c}{$\ce{H_2}$ Singlet $\ce{S_1}$}\\ \hline
    &&\multicolumn{2}{c}{$F^{(1)}(\bm{\theta})$} &&\multicolumn{2}{c}{$F^{(2)}(\bm{\theta})$} \\\hline
    $\mu_C$ && \multicolumn{1}{c}{$N_\mr{meas}$}  & \multicolumn{1}{c}{residuals} && \multicolumn{1}{c}{$N_\mr{meas}$}  & \multicolumn{1}{c}{residuals} \\\hline
    $0.01$ && $\SI{18099}{}$ &$\SI{-0.360693}{}$&&$\SI{1271937}{}$ & $\SI{-0.363527}{}$ \\\hline
    $0.1$ && $\SI{12794}{}$  &$\SI{-0.268809}{}$&&$\SI{986748}{}$ &$\SI{-0.329921}{}$ \\\hline
    $1$ && $\SI{12390}{}$  &$\SI{0.000000}{}$&&$\SI{1308839}{}$ & $\SI{-0.323611}{}$ \\\hline
    $10$ && $\SI{14519}{}$  &$\SI{0.000000}{}$&&$\SI{1264772}{}$  &$\SI{-0.323009}{}$\\\hline
    $100$ && $\SI{32000}{}$  &$\SI{0.000000}{}$&&$\SI{1293515}{}$ &$\SI{-0.322953}{}$ \\\hline 
    \end{tabular}
    \caption{The results of the numerical simulations for calculating the energy of the $\ce{S_1}$ state of $\ce{H_2}$ by the constrained VQD under the constraint $\hat{N}_e = 2$ with various $\mu_C$. 
    Note that we only constrain the number of electrons $\hat{N}_e$ in this simulation while we compute the same energy by constraining $\hat{N}_e$ and $\hat{S}^2$ in the simulation shown in TABLE~\ref{table_H2_mu} of Sec.~\ref{subsec:comparison_mu}.
    $N_\mr{meas}$ represents the number of Pauli measurements necessary to perform the whole optimization process, which is computed as (the number of evaluations of the cost function during the optimization) $\times$ (the number of Pauli operators to be measured to get the value of the cost function once).
    The term ``residuals" represents the difference between the expectation value of the Hamiltonian for the resulting optimized state and that for the exact (desired) state in Hartree (Ha).
    The exact value for this case is $\SI{-0.169901}{Ha}$.
    We perform the constrained VQD for ten different initial values, and all values shown in this table are the average for the ten trials.}
    \label{table_F2_fail}
\end{table}

\section{\label{sec:conclusion}Conclusion}
In this work, we study two cost functions $F^{(1)}(\bm{\theta})$ (Eq.~\eqref{eq:cost_func1}) and $F^{(2)}(\bm{\theta})$ (Eq.~\eqref{eq:cost_func2}) of the constrained VQE/VQD, which can be exploited to compute eigenstates of the Hamiltonian of a given system that reside in the desired symmetry sector. 
Our theoretical analysis revealed that minimization of the cost function $F^{(1)}(\bm{\theta})$ can yield the desired state/energy when the penalty coefficient $\mu_C$ is larger than a certain threshold (Eq.~\eqref{eq:mu_formula}), and we derived a simple and practical formula to estimate it (Eq.~\eqref{eq:estimate_mu}).
On the other hand, we proved that the exact desired state/energy cannot be obtained by minimizing the cost function $F^{(2)}(\bm{\theta})$ and that we obtain completely wrong values in some cases.
To validate these theoretical analyses, we performed several numerical simulations of the constrained VQE/VQD for $\ce{H_2}$ and $\ce{H_4}$ molecules.
Our simulations validated the formula~\eqref{eq:estimate_mu} for $F^{(1)}(\bm{\theta})$ in practical quantum chemistry calculations and indicated that we should estimate the energy gap in the formula as accurately as possible to achieve the fast convergence of the optimization.
Furthermore, we found an explicit example where the desired state/energy is never obtained by using the cost function $F^{(2)}(\bm{\theta})$.
Even though $F^{(2)}(\bm{\theta})$ sometimes shows a better performance than $F^{(1)}(\bm{\theta})$ in terms of the total number of Pauli measurements required for the optimization, $F^{(1)}(\bm{\theta})$ still serves as a better cost function of the constrained VQE/VQD because we can ensure that the target state/energy is obtained as a result of the optimization. 

Our results elucidate the fundamental difference between the performances of the two cost functions $F^{(1)}(\bm{\theta})$ and $F^{(2)}(\bm{\theta})$. 
The inconsistent and heuristic use of the two cost functions is resolved by our results showing the theoretical superiority of $F^{(1)}(\bm{\theta})$ over $F^{(2)}(\bm{\theta})$ and providing the formula~\eqref{eq:estimate_mu} to determine the appropriate penalty coefficient.
The proper choice of the penalty coefficient leads to the fast convergence of the optimization.
Since we assume only the discreteness of the system, our findings apply to general quantum systems and lay the theoretical foundation for exploiting NISQ devices with the constrained VQE/VQD to solve large quantum systems that are classically intractable. 

As future work, it would be intriguing to study the effect of the noise for more general types of noise.
Another interesting direction is to numerically investigate other symmetries that are not treated in this work, such as translation symmetry or point group symmetry.

\begin{acknowledgments}
We thank Ryosuke Imai, Youyang Zhang, and Yohei Ibe for helpful discussions.
We appreciate Kosuke Mitarai and Wataru Mizukami for stimulating discussions and reading the draft of the manuscript.
KK is supported by QunaSys Inc., Mike and Ophelia Lazaridis, and research grants from the NSERC\@.
A part of this work was performed for the Council for Science, Technology and Innovation (CSTI), Cross-ministerial Strategic Innovation Promotion Program (SIP), ``Photonics and Quantum Technology for Society 5.0'' (Funding agency: QST).
\end{acknowledgments}

\bibliography{bibliography}
\end{document}